\documentclass[a4paper]{jpconf}
\usepackage{graphicx}
\usepackage{amsmath,amssymb}
\begin{document}
\title{Double-$\it q$ Order in a Frustrated Random Spin System}

\author{Ryo Tamura and Naoki Kawashima}
\vspace{5pt}
\address{Institute for Solid State Physics, University of Tokyo, 5-1-5 Kashiwa-no-ha, Kashiwa, Chiba 277-8581, Japan}

\ead{r.tamura@issp.u-tokyo.ac.jp}

\begin{abstract}
We use the three-dimensional Heisenberg model with site randomness as an effective model of the compound Sr(Fe$_{1-x}$Mn$_x$)O$_2$.
The model consists of two types of ions that correspond to Fe and Mn ions.
The nearest-neighbor interactions in the $ab$-plane are antiferromagnetic.
The nearest-neighbor interactions along the $c$-axis between Fe ions are assumed to be antiferromagnetic,
whereas other interactions are assumed to be ferromagnetic.
From Monte Carlo simulations,
we confirm the existence of the double-$\boldsymbol{q}$ ordered phase characterized by two wave numbers, $(\pi\pi\pi)$ and $(\pi\pi0)$.
We also identify the spin ordering pattern in the double-$\boldsymbol{q}$ ordered phase.

\end{abstract}


\section{Introduction}

Infinite-layer iron oxide SrFeO$_2$ exhibits a N$\acute{\text{e}}$el transition to the G-type antiferromagnetic ordered phase at $T_N=473$K\cite{Tsujimoto2007,Xiang2008}.
The transition temperature decreases as Mn$^{2+}$ ions are substituted for Fe$^{2+}$ ions\cite{Kageyama2010}.
In the compound Sr(Fe$_{1-x}$Mn$_x$)O$_2$,
Fe and Mn ions are magnetic and placed randomly on a tetragonal lattice.
In neutron scattering measurements on Sr(Fe$_{0.7}$Mn$_{0.3}$)O$_2$,
magnetic peaks have been observed to develop at two distinct wave vectors $\boldsymbol{q}=(\pi\pi\pi)$ and $(\pi\pi0)$. 
These orders coexist at low temperatures\cite{Kageyama2010}.
However,
the spin ordering pattern in the low-temperature phase is not known.

The site-random model has been used as an effective model of the random magnets.
The site-random model consists of two types of ions labeled A and B.
The ions are placed randomly on the lattice.
The interactions of each bond are set depending on the combination of ions.
In previous works that employed the site-random model\cite{Hukushima1997,Nielsen1996,Matsubara1996,Bekhechi2004,Beath2006},
all interactions depended on the arrangement of A and B ions:
the interactions between A ions were ferromagnetic,
whereas other interactions were antiferromagnetic.
In the case of the Ising spin model on a simple cubic lattice\cite{Hukushima1997},
it was found through Monte Carlo simulations that the phase diagram lacks both a mixed ordered phase and spin-glass phase.
In the Heisenberg spin model\cite{Nielsen1996,Matsubara1996,Bekhechi2004,Beath2006},
as in the Ising spin case,
there is also no spin-glass phase in the phase diagram.
However,
the ferromagnetic and antiferromagnetic mixed ordered phase exists at low temperatures.
In this mixed ordered phase,
the peaks of the structure factor develop at wave vectors $\boldsymbol{q}=(000)$ and $(\pi\pi\pi)$.
Moreover,
the ferromagnetic and antiferromagnetic ordering vectors are mutually perpendicular.
For this model,
the occurrence of a two-step phase transition has been reported:
one is the transition from the paramagnetic phase to the ferromagnetic (antiferromagnetic) ordered phase,
the other is the transition from the ferromagnetic (antiferromagnetic) ordered phase to the mixed ordered phase.
The universality class at each phase transition has been suggested to belong to the three-dimensional Heisenberg universality class\cite{Beath2006}.

In the previous works that used the site-random Heisenberg model\cite{Nielsen1996,Matsubara1996,Bekhechi2004,Beath2006},
$\boldsymbol{q}=(000)$ and $(\pi\pi\pi)$ orderings have been found.
However,
since the magnetic peak positions are $\boldsymbol{q}=(\pi\pi\pi)$ and $(\pi\pi0)$ in Sr(Fe$_{0.7}$Mn$_{0.3}$)O$_2$,
the spin ordering pattern is different from that observed previously.
Thus,
our current aim is to clarify the spin ordering pattern in the low-temperature phase of Sr(Fe$_{1-x}$Mn$_{x}$)O$_2$.


\section{Model}
To introduce suitable randomness into Sr(Fe$_{1-x}$Mn$_{x}$)O$_2$,
we investigate a classical Heisenberg model with site-random interlayer couplings on a simple cubic lattice:
\begin{align}
\mathcal{H}=-J \sum_{\langle i,j \rangle_\text{$ab$-plane}} \boldsymbol{s}_i \cdot \boldsymbol{s}_j - \sum_{\langle i,j \rangle_\text{$c$-axis}} J_{ij} \boldsymbol{s}_i \cdot \boldsymbol{s}_j, \ \ \ \ \ (|J_{ij}|=|J|).
\end{align}
The first term of the Hamiltonian denotes the uniform antiferromagnetic nearest-neighbor interaction ($J<0$) in the $ab$-plane.
The sign of the nearest-neighbor coupling $J_{ij}$ in the second term depends on the arrangement of two ions along the $c$-axis;
the interactions between Fe ions are assumed to be antiferromagnetic ($J_{ij}<0$),
whereas other interactions are assumed to be ferromagnetic ($J_{ij}>0$).
For simplicity,
we further assume that the absolute values of all interactions are the same
and that the spin lengths of Fe and Mn ions are the same.
Hereinafter, we set $|J|=1$ and denote the concentration of Mn ions by $p_\text{Mn}$.
From the above assumptions regarding the interactions,
at $p_\text{Mn}=0$
where all magnetic ions are Fe ions,
the system exhibits a N$\acute{\text{e}}$el transition to the G-type antiferromagnetic ordered phase characterized by $\boldsymbol{q}=(\pi\pi\pi)$,
because all interactions along the $c$-axis are antiferromagnetic.
In contrast,
at $p_\text{Mn}=1$
where all magnetic ions are Mn ions,
the C-type antiferromagnetic ordered phase characterized by $\boldsymbol{q}=(\pi\pi0)$ becomes stable at low temperatures,
because all interactions along the $c$-axis are ferromagnetic.
For $0<p_\text{Mn}<1$,
the frustration is caused by the random ferromagnetic and antiferromagnetic interactions along the $c$-axis direction.


\section{Simulation Results}

We use Monte Carlo simulations based on the standard heat-bath method to the model on an $N=L \times L \times L$ simple cubic lattice with a periodic boundary condition.
Before starting the simulations,
we set $p_\text{Mn}$ so that the number of Mn ions is an integer,
and we set random configurations of Fe and Mn ions according to $p_\text{Mn}$.
In this paper,
$p_\text{Mn}$ is set to 0.1875.
For this value of $p_\text{Mn}$,
there are more antiferromagnetic couplings than ferromagnetic couplings along the $c$-axis.
We prepare 64 samples,
and each run contains $10^6$--$10^7$ Monte Carlo steps per spin at each temperature.

\begin{figure}[t]
\begin{center}
\includegraphics[trim=0mm 0mm 0mm 0mm ,scale=0.35, angle=270]{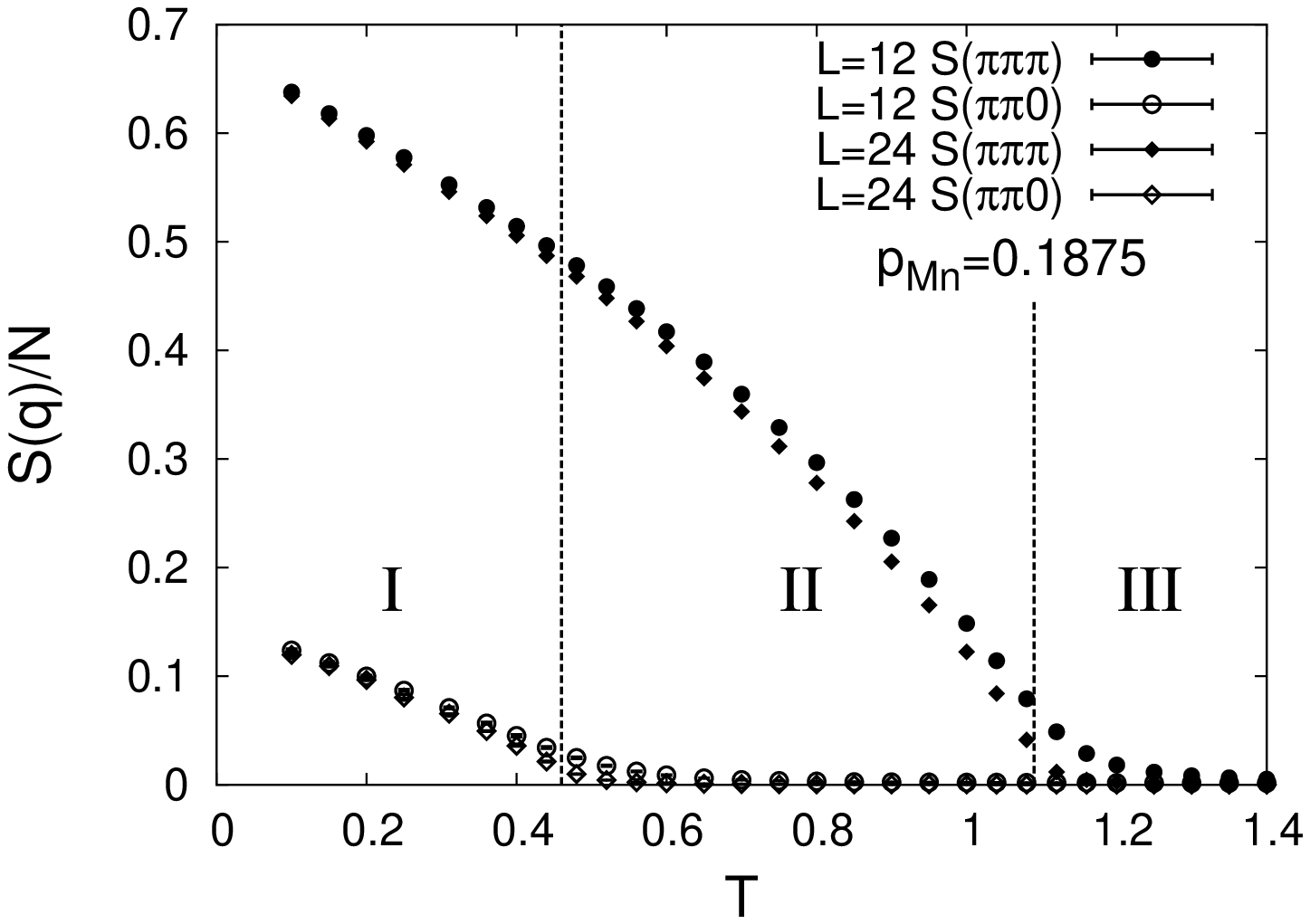} 
\end{center}
\caption{\label{fig:Sq} 
Temperature dependence of the structure factor at $\boldsymbol{q}=(\pi\pi\pi)$ and $(\pi\pi0)$ when the lattice size is $L=12$ and 24.
Phases denoted by I, II, and III are the double-$\boldsymbol{q}$ ordered phase, the G-type antiferromagnetic ordered phase, and the paramagnetic phase, respectively.
}
\end{figure}

The temperature dependence of the structure factor $S (\boldsymbol{q})$ at $\boldsymbol{q}=(\pi\pi\pi)$ and $(\pi\pi0)$ is shown in Figure~\ref{fig:Sq}.
The structure factor $S(\boldsymbol{q})$ is defined by
\begin{align}
S (\boldsymbol{q}) &= \frac{1}{N} \sum_{i,j} \langle \boldsymbol{s}_i \cdot \boldsymbol{s}_j \rangle 
e^{i \boldsymbol{q} \cdot (\boldsymbol{r}_i-\boldsymbol{r}_j)},
\end{align}
where $\langle \cdots \rangle$ indicates the thermal average.
$S(\pi\pi\pi)$ and $S(\pi\pi0)$ increase at $T\cong 1.09$ and 0.46, respectively.
From Figure~\ref{fig:Sq},
it is clear that there are three types of phases in this model.
In the low-temperature phase (phase I),
the peaks of the structure factor develop at two distinct wave vectors $\boldsymbol{q}=(\pi\pi\pi)$ and $(\pi\pi0)$.
Accordingly,
this phase is called a double-$\boldsymbol{q}$ ordered phase.
In the intermediate-temperature phase (phase II),
since the structure factor has a peak at $\boldsymbol{q}=(\pi\pi\pi)$,
this phase is the G-type antiferromagnetic ordered phase.
The high-temperature phase (phase III) is the paramagnetic phase.
The transition temperature from the G-type antiferromagnetic ordered phase to the paramagnetic phase is lower than the N$\acute{\text{e}}$el temperature of the pure system $T_N\cong 1.44$\cite{Chen1993}.
From the results of the structure factor,
we conclude that the double-$\boldsymbol{q}$ ordered phase characterized by the wave vectors $\boldsymbol{q}=(\pi\pi\pi)$ and $(\pi\pi0)$ exists in the model. 
The positions of the magnetic peaks for the model are the same as those observed for Sr(Fe$_{1-x}$Mn$_x$)O$_2$.

@

\begin{figure}[t]
\begin{center}
\includegraphics[trim=0mm 0mm 0mm 0mm ,scale=0.35, angle=270]{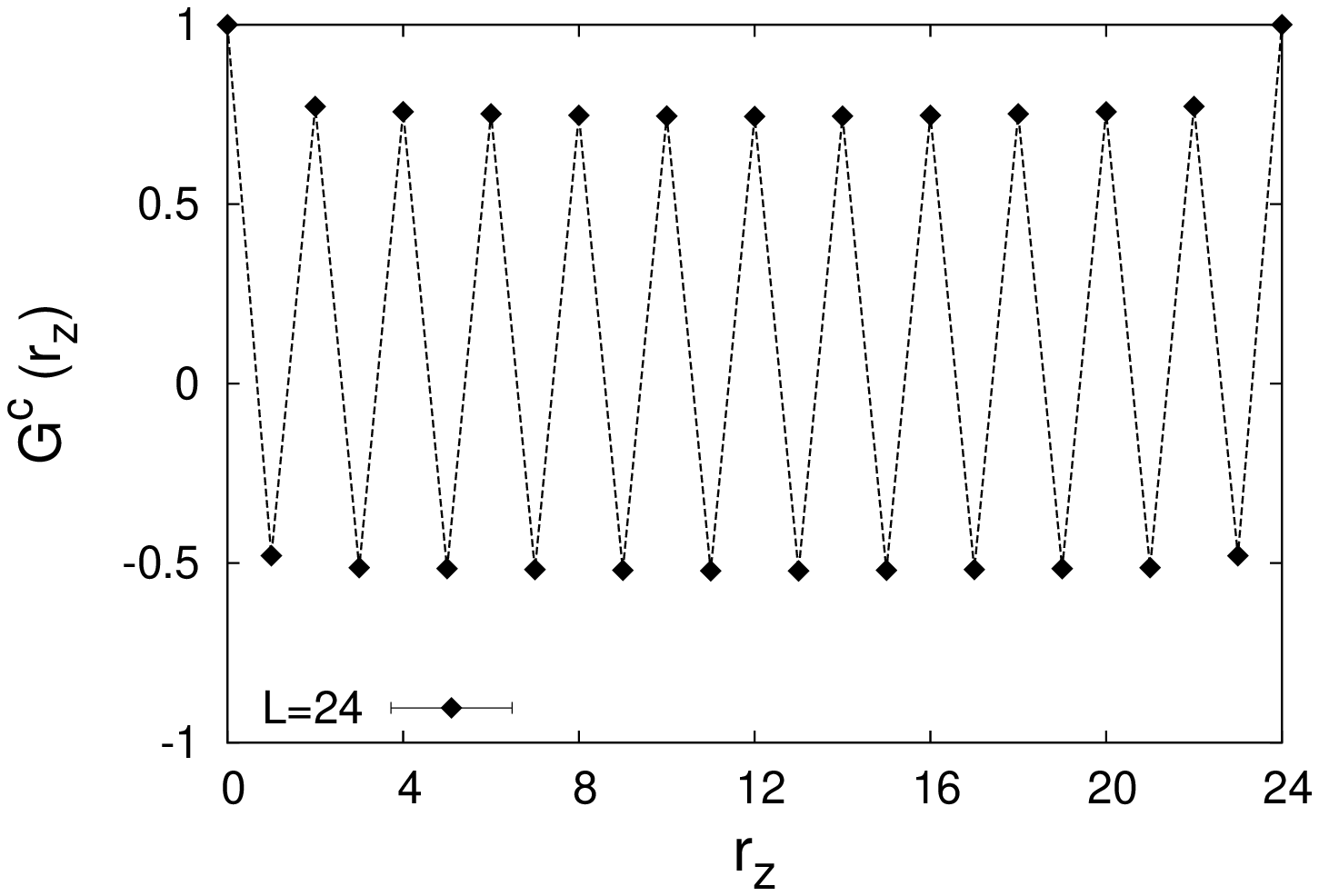} 
\end{center}
\caption{\label{fig:corr_z} 
Correlation function along the $c$-axis for $L=24$ at $T=0.1$.
}
\end{figure}

To identify the spin ordering pattern in the double-$\boldsymbol{q}$ ordered phase,
we calculate the correlation function along the $c$-axis between two spins.
The correlation function along the $c$-axis is given by
\begin{align}
G^c(r_z)=\frac{1}{N} \sum_i \langle \boldsymbol{s} (\boldsymbol{r}_i) \cdot \boldsymbol{s} (\boldsymbol{r}_i+ r_z \boldsymbol{e}_z) \rangle,
\end{align}
where $r_z$ is the distance between two spins, 
and $\boldsymbol{e}_z$ is unit vector of the $c$-axis direction.
Figure~\ref{fig:corr_z}  shows dependence of the correlation function on distance $r_z$ for $L=24$ at $T=0.1$.
Since the spin correlation is ferromagnetic for an even number of $r_z$,
all spins in odd (even)-numbered layers are co-linear;
The N$\acute{\text{e}}$el-like configuration of each layer is stacked in parallel to other layers.
Furthermore,
the spin correlation is weakly coupled antiferromagnetic for an odd-number of $r_z$.
This suggests that the angle between nearest-neighbor spins along the $c$-axis is close to $2\pi/3$ at $T=0.1$.

\begin{figure}[t]
\begin{center}
\includegraphics[trim=0mm 0mm 0mm 0mm ,scale=0.35, angle=270]{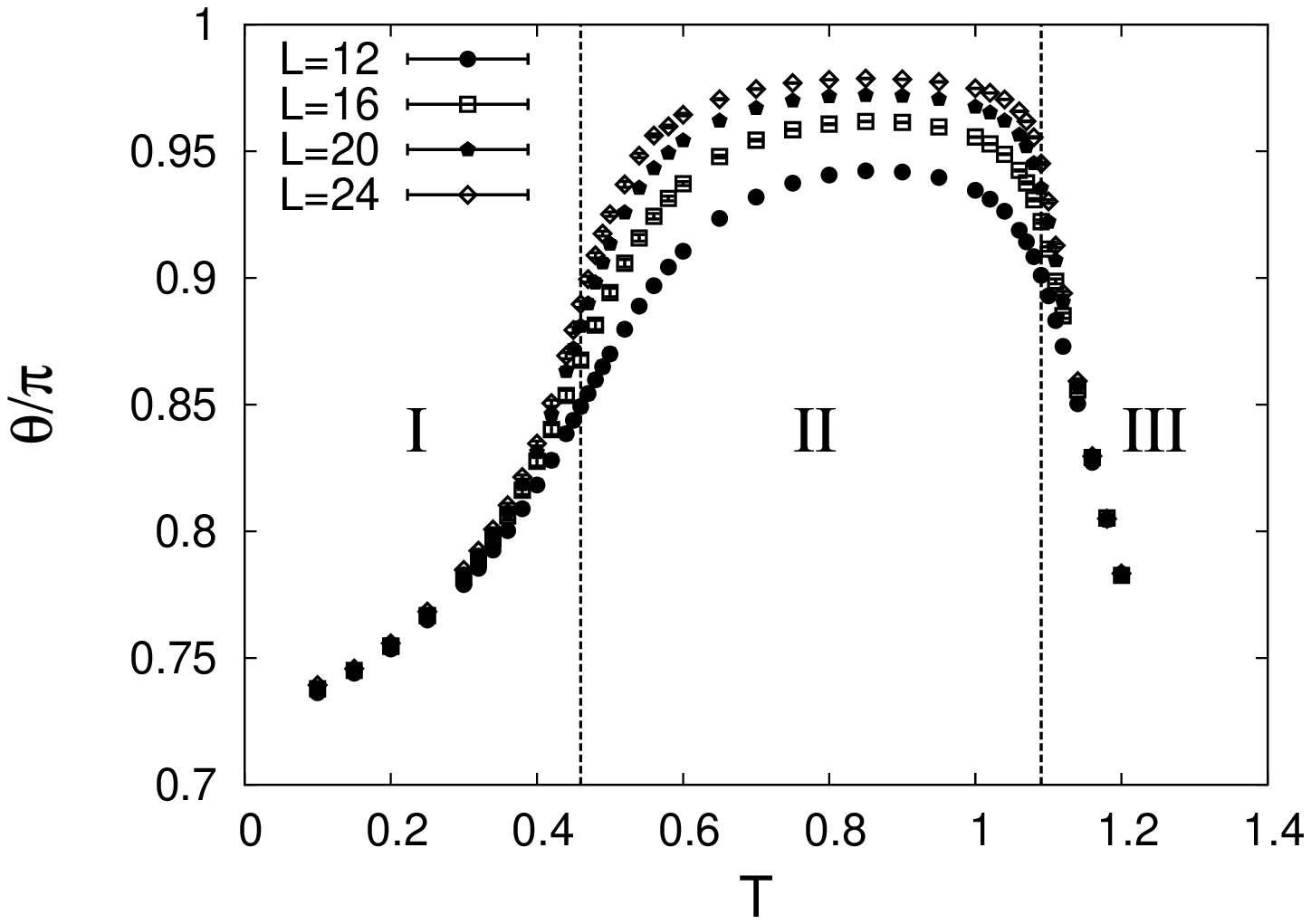} 
\end{center}
\caption{\label{fig:angle} 
Temperature dependence of the angle $\theta$ between the staggered magnetization of an odd-numbered layer and that of an even-numbered layer.
}
\end{figure}

To investigate the temperature dependence of spin ordering pattern,
we calculate the angle $\theta$ between the staggered magnetization of an odd-numbered layer, $\boldsymbol{M}_o$, and that of an even-numbered layer, $\boldsymbol{M}_e$.
$\boldsymbol{M}_o$ and $\boldsymbol{M}_e$ are given by
\begin{align}
\boldsymbol{M}_o = \frac{2}{N} \sum_{\substack{ \boldsymbol{r} \in \text{odd-numbered} \\ \text{layer}}} e^{i \boldsymbol{q} \cdot \boldsymbol{r}} \boldsymbol{s} (\boldsymbol{r}), \\
\boldsymbol{M}_e = \frac{2}{N} \sum_{\substack{ \boldsymbol{r} \in \text{even-numbered} \\ \text{layer}}} e^{i \boldsymbol{q} \cdot \boldsymbol{r}} \boldsymbol{s} (\boldsymbol{r}),
\end{align}
where $\boldsymbol{q}=(\pi\pi0)$.
We define the angle $\theta$ between $\boldsymbol{M}_o$ and $\boldsymbol{M}_e$ as follows:
\begin{align}
\cos \theta = \left\langle \frac{\boldsymbol{M}_o \cdot \boldsymbol{M}_e}{|\boldsymbol{M}_o||\boldsymbol{M}_e|} \right\rangle.
\end{align}
Figure~\ref{fig:angle} shows the temperature dependence of $\theta$.
At the higher-temperature transition point,
the value of $\theta/\pi$ approaches $1$
and the $(\pi\pi\pi)$ order develops.
At the lower-temperature transition point,
the value of $\theta/\pi$ deviates from $1$.
In phase I,
the value of $\theta$ depends on the temperature and monotonically decreases as the temperature decreases.
Thus,
the spin configuration characterized by $\theta$ changes depending on the temperature in phase I.
At a temperature of zero,
the value of $\theta/\pi$ appears to converge to $0.73$.
From the results,
we find that the lower-temperature phase transition is caused by $\theta/\pi$ deviating from 1.

Let us consider the translational symmetry and the spin rotation symmetry in each phase.
The intermediate-temperature phase is translationally symmetric with the O(3) spin rotation symmetry broken down to the U(1).
In the double-$\boldsymbol{q}$ ordered phase,
both the translational symmetry and the U(1) spin rotation symmetry are broken.
Thus,
we conclude that the transition to the double-$\boldsymbol{q}$ ordered phase is characterized by the breaking of the U(1) spin rotation symmetry.


\section{Discussion}
We discuss dependence of physical properties on the concentration of Mn ions $p_\text{Mn}$.
In this paper,
we have fixed $p_\text{Mn}=0.1875$.
In this case, 
there are more antiferromagnetic couplings than ferromagnetic couplings along the $c$-axis, 
and the intermediate G-type antiferromagnetic ordered phase exists in a broad temperature range ($0.46 \lesssim T \lesssim 1.09$).
At $p_{\text{Mn}} \cong 0.3$,
the number of antiferromagnetic couplings and ferromagnetic couplings along the $c$-axis become nearly the same.
Then,
$(\pi\pi\pi)$ order and $(\pi\pi0)$ order develop at nearly the same temperature,
and the intermediate-temperature phase exists in a very narrow temperature range.
Furthermore,
for $p_{\text{Mn}} > 0.3$,
the intermediate C-type antiferromagnetic ordered phase exists,
and the transition to the double-$\boldsymbol{q}$ ordered phase occurs at low temperature.
In future work, we will attempt to clarify the details of the phase diagram of the temperature versus the concentration of Mn ions.


\section{Conclusion}

We have used the three-dimensional Heisenberg model with site-random interlayer couplings as an effective model of the compound Sr(Fe$_{1-x}$Mn$_x$)O$_2$.
The model consists of two types of ions that correspond to Fe and Mn ions.
The interactions have been assumed to be antiferromagnetic in the $ab$-plane.
The interactions along the $c$-axis between Fe ions have been assumed to be antiferromagnetic,
whereas other interactions have been assumed to be ferromagnetic.
From Monte Carlo simulations,
we have confirmed the existence of the double-$\boldsymbol{q}$ ordered phase characterized by the wave vectors $\boldsymbol{q}=(\pi\pi\pi)$ and $(\pi\pi0)$.
The positions of the magnetic peaks for the model are the same as those observed for Sr(Fe$_{1-x}$Mn$_{x}$)O$_2$.
We have also identified the spin ordering pattern in the double-$\boldsymbol{q}$ ordered phase.
In the double-$\boldsymbol{q}$ ordered phase,
all spins in odd (even)-numbered layers are co-linear.
Furthermore,
the staggered magnetization of an odd-numbered layer and that of an even-numbered layer are not parallel,
and the angle between them changes depending on the temperature.
We conclude that the transition to the double-$\boldsymbol{q}$ ordered phase is characterized by the breaking of the U(1) spin rotation symmetry.


\ack{
We would like to thank Takafumi Suzuki, Shu Tanaka, Liis Seinberg, Takafumi Yamamoto, and Hiroshi Kageyama for useful comments and discussions.
R.T. acknowledges support from the Global COE Program ``The Physical Sciences Frontier'' of the Ministry of Education, Culture, Sports, Science and Technology (MEXT), Japan.
This work is financially supported by Grants-in-Aid for Scientific Research (B) (22340111) and for Scientific Research on Priority Areas ``Novel States of Matter Induced by Frustration'' (19052004) from MEXT, Japan, and by Next Generation Supercomputing Project, Nanoscience Program, MEXT, Japan.
The computation in this work is executed on computers at the Supercomputer Center, Institute for Solid State Physics, University of Tokyo. }


\end{document}